# TaiBai: A fully programmable brain-inspired processor with topology-aware efficiency

Qianpeng Li, Yu Song, Xin Liu, Wenna Song, Boshi Zhao, Zhichao Wang, Aoxin Chen,
Tielin Zhang* and Liang Chen*

*Abstract*—Brain-inspired computing has emerged as a promising paradigm to overcome the energy-efficiency limitations of conventional intelligent systems by emulating the brain's partitioned architecture and event-driven sparse computation. However, existing brain-inspired chips often suffer from rigid network topology constraints and limited neuronal programmability, hindering their adaptability. To address these challenges, we present TaiBai, an event-driven, programmable many-core brain-inspired processor that leverages temporal and spatial spike sparsity to minimize bandwidth and computational overhead. TaiBai chip contains three key features: First, a brain-inspired hierarchical topology encoding scheme is designed to flexibly support arbitrary network architectures while slashing storage overhead for large-scale networks; Second, a multi-granularity instruction set enables programmability of brain-like spiking neuron or synapses with various dynamics and on-chip learning rules; Third, a co-designed compiler stack optimizes task mapping and resource allocation. After evaluating across various tasks, such as speech recognition, ECG classification, and cross-day brain-computer interface decoding, we found spiking neural networks embedded on the TaiBai chip could achieve more than 200 times higher energy efficiency than a standard NVIDIA RTX 3090 GPU at a comparable accuracy. These results demonstrated its high potentation as a scalable, programmable, and ultra-efficient solution for both multi-scale brain simulation and brain-inspired computation.

*Index Terms*—Brain-inspired computing chip, spiking neural network, fully programmable, network topology representation.

## I. INTRODUCTION

BUILDING a vast digital simulation of the brain could transform neuroscience and medicine and reveal new ways of making more powerful computers [1]. The human brain has nearly 100 billion neurons and $10^{15}$ synapses. It operates the most complex, large, efficient and robust neural system yet known, with a power consumption of ~20W and extraordinary task-processing capabilities. Even with the most advanced digital supercomputers available, it takes an enormous amount of energy and time to simulate neural networks of the same size [2].

As a result, brain-inspired platforms are being explored by drawing on brain organization and information processing patterns, from the device to the circuit to the architectural level [3], various brain-inspired chips are designed and power overheads are reduced. TrueNorth [4], SpiNNaker [5], Loihi [6], and Tianjic [7] have demonstrated superior low-energy computing benefits for tasks such as speech recognition, target tracking, and control decision-making.

Physiological studies in recent years have established a causal relationship between heterogeneous dendritic axons and the firing properties of neurons [8], neuronal heterogeneity plays an important role in determining the function of neural circuits [9]-[11] and determines the frequency of the neural spike signal [12]. Neurons have numerous specific connections to each other [13], and different connections support different neural circuit functions [14]. The brain has the capacity to achieve complex learning and memory functions in a variety of environments through diverse connection modes. Spiking neural networks (SNNs) has been shown to achieve energy-efficient artificial intelligence by drawing on biological mechanisms such as the brain's heterogeneous neurons and network structures, maintaining biological consistency at multiple scales, including axon-dendrite structure, neuronal dynamics, and neuronal connectivity. SNNs exploit these heterogeneities to provide the ability to capture multiscale temporal features [15], [16], improving the accuracy and robustness of models [17]-[19].

Despite these promising biological insights, current brain-inspired hardware architectures suffer from significant limitations. Existing brain-inspired platforms face substantial limitations in programmable capabilities and network topology representations. Current hardware architectures generally support limited neuron models, such as the leaky integrate-and-fire (LIF), restricting their ability to accurately replicate diverse and complex biological neuron behaviors. Additionally, these platforms often implement fixed or minimally adaptable learning algorithms, significantly constraining their applicability to diverse and dynamic computational scenarios. Network topology representation in

Qianpeng Li, Zhichao Wang, Aoxin Chen and Liang Chen are with Institute of Automation, Chinese Academy of Sciences (CASIA), Beijing 100190, China, and also with School of Artificial Intelligence, University of Chinese Academy of Sciences (UCAS), Beijing 100049, China. Yu Song is with Institute of Automation, Chinese Academy of Sciences (CASIA), Beijing 100190, China and also with Center for Excellence in Brain Science and Intelligence Technology, Chinese Academy of Sciences, Shanghai 200031, China. Xin Liu, Wenna Song are with Institute of Automation, Chinese Academy of Sciences (CASIA), Beijing 100190, China. Boshi Zhao is with Center for Excellence in Brain Science and Intelligence Technology, Chinese Academy of Sciences, Shanghai 200031, China. Tielin Zhang is with the Center for Excellence in Brain Science and Intelligence Technology, Chinese Academy of Sciences, Shanghai 200031, China, School of Artificial Intelligence, University of Chinese Academy of Sciences, Beijing 100049, China, and also with State Key Laboratory of Brain Cognition and Brain-inspired Intelligence Technology, Shanghai 200031, China.

The corresponding authors are Tielin Zhang and Liang Chen.

current brain-inspired platforms further exacerbates these limitations. Conventional approaches, including crossbars and fan-in/fan-out tables, encounter significant scalability challenges and suffer from inefficient memory usage, particularly in convolutional networks where weight sharing is prevalent. This redundancy in representation reduces computational efficiency and increases resource demands, severely restricting the deployment and execution of large-scale and sophisticated neural networks.

To overcome these limitations, we introduce **TaiBai**, a fully programmable brain-inspired computing chip designed from the ground up to encapsulate essential biological principles of neuronal heterogeneity, synaptic heterogeneity, flexible network topologies, and on-chip synaptic plasticity. TaiBai integrates an innovative event-driven architecture with a hierarchical and compact network representation strategy, significantly enhancing scalability and resource efficiency. TaiBai uses a Turing-complete brain-inspired instruction set that enables the flexible implementation of diverse neuron and synapse models, as well as multiple on-chip learning algorithms. The chip's event-driven scheduler dynamically manages computations across 132 Cortical Column (CC) cores interconnected through a 2D mesh Network-on-Chip (NoC), optimizing parallelism and reducing redundant operations. Supported by an end-to-end compiler stack optimized for brain-inspired workloads, TaiBai substantially streamlines model deployment, fully leveraging hardware potential. Comprehensive evaluations across tasks such as speech recognition, electrocardiogram (ECG) classification, and brain-computer interface (BCI) decoding demonstrate TaiBai's remarkable improvements in energy efficiency, power consumption, and computational accuracy, positioning it as a highly versatile and powerful solution for next-generation brain-inspired computing.

## II. BACKGROUND AND MOTIVATION

### A. Spiking Neural Network

Biological LIF neurons are widely used in brain simulations and SNNs, their membrane potential integrates incoming spike currents until a threshold is reached, at which point an output spike is generated and the potential resets. When there is a spike input, the membrane potential increases by weight, otherwise the membrane potential will decay over time. If the membrane potential is greater than or equal to threshold $V_{th}$, it will be reset to zero and a spike will be fired. The equations of the model are shown as (1)-(3),

$$I(t) = \sum_i x_i(t) w_i, \quad (1)$$
$$v(t) = \tau v(t-1) + I(t), \quad (2)$$
$$\begin{cases} v(t) = 0, s(t) = 1 & if\ v(t) \geq V_{th} \\ v(t) = v(t), s(t) = 0 & Otherwise \end{cases}, \quad (3)$$

$I(t)$ is the accumulated current of the neuron at timestep $t$, $w_i$ is the synaptic weight with the $i$-th neuron, $v(t)$ is the membrane potential of neuron at timestep $t$, $\tau$ is the decay factor, $s(t)$ is the output spike of the neuron at timestep $t$.

Spike-timing-dependent plasticity (STDP) [20] is an unsupervised, Hebbian-style learning rule that adjusts each synapse according to the precise time difference ($\Delta t$) between presynaptic and postsynaptic spikes: causal pairs ($\Delta t > 0$) potentiate the weight, whereas acausal pairs ($\Delta t < 0$) depress it. Because the update depends only on locally available spike times, STDP can run online in a fully event-driven manner, making it well suited for low-power, on-chip adaptation.

Spatio-temporal backpropagation (STBP) [21], in contrast, extends global gradient-descent learning to spiking neural networks. It replaces the non-differentiable firing operation with a surrogate gradient—typically a smooth sigmoid or piecewise-linear proxy—so that standard back-propagation through time (BPTT) can be applied across unfolded timesteps. This approach attains state-of-the-art accuracy on image, speech, and reinforcement-learning benchmarks, but it incurs higher memory and computational overhead than local rules such as STDP.

### B. Brain-inspired Computing Processor

Mainstream brain-inspired computing processors such as TrueNorth [4], Loihi [6], Loihi2 [22], and Tianjic [7] abandon the architecture of traditional ANN accelerators to accelerate a single network layer in the form of an array, and instead adopt a global tiling mode of interconnected neuron cores based on the NoC. The neuron core includes multiple neurons, synaptic connections, and synaptic weights. When a neuron fires a spike, the spike information can be transmitted to the processing core where the postsynaptic neuron is located through the NoC, thereby triggering the update of the neuron membrane potential. This structure avoids the power consumption and delay overhead caused by the ANN accelerator using external DRAM to load and recycle data by placing neurons, synapses, and network topology entirely on the processor.

### C. Limitations in Programmability

Most current brain-inspired processors are only partially programmable, being tailored to one or two neuron models or fixed learning rules, which sharply limits their adaptability to diverse tasks and environments. Brain-inspired platforms such as TrueNorth, Neurogrid [23], Loihi, Tianjic, and some new device-based chips only support LIF neuron models, making it difficult to simulate complex neuron models. Flexon [24] and FlexLearn [25] consolidate sub-rules for neural models and synaptic learning algorithms respectively, and can only inefficiently simulate other biological behavior by combining sub-rules. PAICORE [26] only provides the local on-chip learning algorithm, with limited support for multi-layer network learning. Loihi2 only provides a three-factor based synaptic on-chip learning method, which is difficult to adapt to richer and more flexible learning algorithms. Darwin3 [27], based on a domain-specific instruction set architecture and specific data paths, provides limited programmable capabilities for neural models. It only supports bionic local learning algorithms and has limitations in model scalability. SpiNNaker provides fully programmable capabilities based on Arm processors, but relies on limited biological brain mechanisms and is not dominant in terms of hardware resource in utilization and energy efficiency, limiting its potential for use resource-constrained and high energy-efficient scenarios.

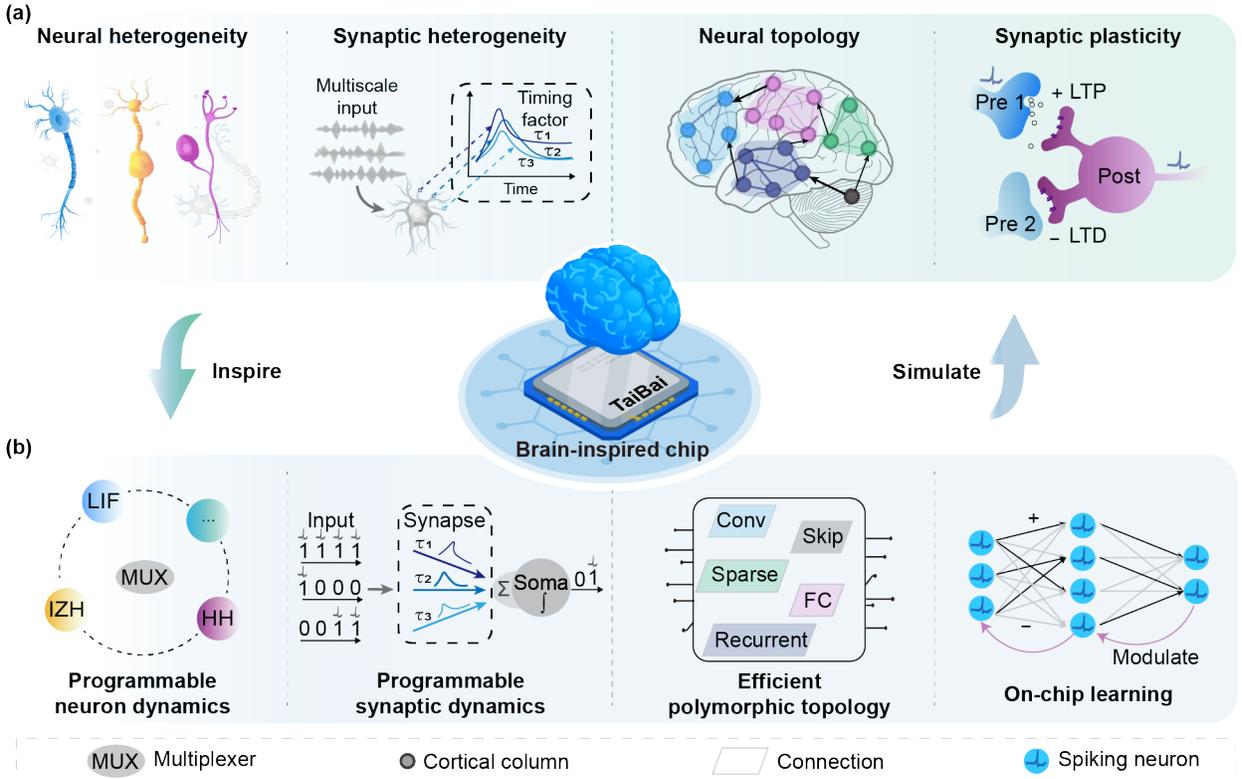

Fig. 1. The chip is inspired by (a) the key characteristics of brain structure to enhance (b) multiple programmable capabilities and efficient representation of any network topologies.

*D. Limitations in Network Topology Representation*

In terms of network topology representation, each core of the brain-inspired chip stores and executes a partial SNN model [28], [29] and requires dedicated logic to implement the complex connection structure between neurons [30], [31]. The network topology representation scheme affects the on-chip memory requirements on the one hand, and on the other hand it also affects the speed at which the destination core can parse spike events, thereby affecting the chip's operating efficiency.

TrueNorth, Tianjic uses a configurable crossbar to represent the topology, a structure with limited fan-in and fan-out, leading to difficulties in deploying large network. This approach suffers from the difficulty of deploying large network models due to the repeated storage of weights when deploying convolutional topologies. Loihi, Darwin3, and LAMR [32] use fan-in and fan-out tables to represent network topologies, which are more flexible than crossbars. However, these existing designs using tables have difficulties in fully supporting common network topologies such as convolutional, fully connected, sparse and skip connections. Moreover, the existing table design exhibits limitations in coping with topologies characterized by highly multiplexed weights, such as convolution. This hinders effective mining and utilization of potentially shared information in these topologies, resulting in redundancy in information processing and reduced resource utilization.

*E. Opportunities*

Recent efforts have explored brain-inspired computing implementations based on analog circuits [33], [34] and emerging device technologies such as memristors [35], [36]. While these approaches hold promise in terms of device density, power efficiency, and native emulation of biological synaptic dynamics, they face substantial obstacles to practical large-scale deployment. These challenges primarily stem from process immaturity, limited compatibility with existing CMOS fabrication techniques, significant device variability, and high manufacturing costs [29]. Consequently, digital brain-inspired solutions remain the most feasible pathway toward scalable and commercially viable brain-inspired computing.

Nonetheless, existing digital brain-inspired chips continue to exhibit notable constraints, particularly regarding their programmability and their ability to flexibly represent diverse neuron dynamics models and complex network topologies. Fig. 1 schematically links the biological inspirations that motivate our work—spanning neuron models, synaptic mechanisms, learning algorithms, and network structures—to the corresponding architectural abstractions required on silicon. As brain-inspired computing research advances rapidly—often requiring iterative refinements of neuron models, synaptic mechanisms, learning algorithms, and network structures—there is an urgent demand for a fully programmable, highly flexible chip architecture. Such a platform must efficiently support diverse computational paradigms, facilitate rapid hardware-software co-development, and adapt seamlessly to evolving scientific and application-oriented demands.

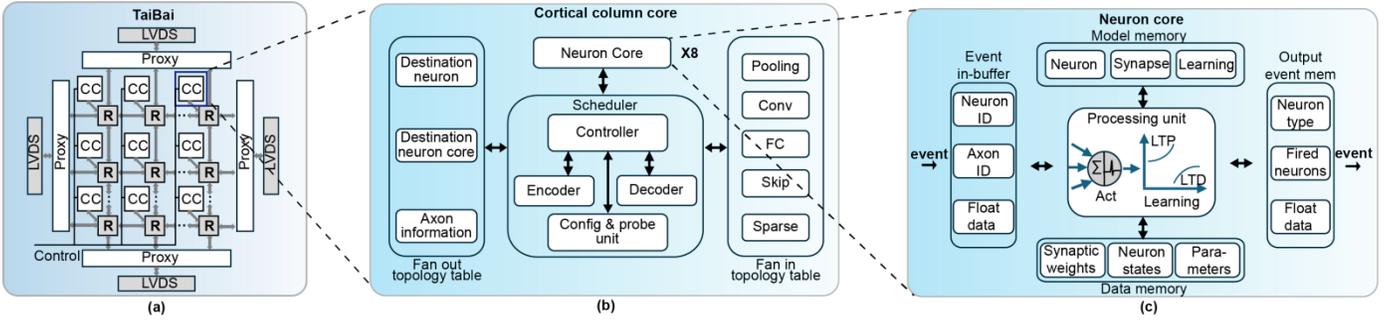

Fig. 2. Overview of TaiBai. (a) The TaiBai chip, (b) cortical column core and (c) neuron core.

Therefore, substantial opportunities exist to develop brain-inspired processors that bridge the gap between biological realism, computational flexibility, and hardware practicality, enabling transformative advances in brain-inspired intelligence research.

## III. THE ARCHITECTURE OF TAIBAI

In this section we detail the architectural foundations of TaiBai, moving from the high-level chip organisation down to the key abstractions that enable full programmability and various topologies. Section III-A presents an overview of the TaiBai chip. Section III-B introduces the brain-inspired instruction set that exposes rich neuron, synapse, and plasticity primitives to software while remaining hardware-efficient. Section III-C describes the router with hybrid-mode routing to balance congestion and latency across the 2-D mesh NoC. Finally, Section III-D describes the network-topology representation scheme, a hierarchical fan-in/out table that supports convolutional, fully connected, sparse, and skip connection without weight replication. Together, these components form the basis for TaiBai's scalable, fully programmable brain-inspired computing platform.

### A. Overview of TaiBai

We developed a brain-inspired computing chip, TaiBai, to solve the problems of current brain-inspired chips and efficiently run SNN models. The primary challenge confronting TaiBai design is the exploitation of model sparsity, with due consideration for model flexibility, the storage overhead of topological representations, and the computational efficiency of the model. The overall architecture of TaiBai is illustrated in Fig. 2(a), comprising 11*12 Cortical column Core(CC) arrays that communicate via a 2D mesh NoC. Routers with hybrid mode routing strategies achieve efficient transmission of information. The chip is surrounded by proxy units and high-speed interfaces for direct chip expansion. The control signal is used to control the operating state of the chip, ensuring the consistency and repeatability of the network model calculation process. Similar to how cortical columns act as the fundamental functional units of the cerebral cortex, the CC (Fig. 2(b)) also serves as the basic functional unit in TaiBai for running SNNs. CC can efficiently encode the network, merge and decode upstream spike information, encode downstream spike information, and schedule the neuron cores (NCs) for model computation. Inspired by biological neurons the NC based on the brain-inspired instruction set can flexibly implement various models (Fig. 2(c)).

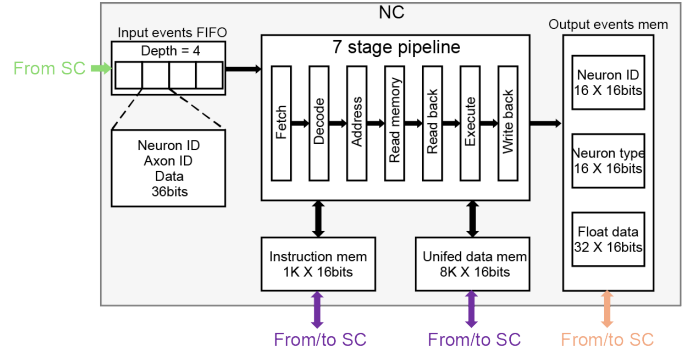

Fig. 3. The architecture of neuron core.

### B. Neuron core with Brain-inspired instruction set

In order to implement different models and on-chip learning algorithms, we designed the NC based on a Turing-complete instruction set. The NC supports two data formats: 16-bit floating point (FP16) and 16-bit integer (INT16). This meets the needs of neuroscience for high computational accuracy and complex model modeling. We designed five special instructions (first five in Table I) for brain-inspired computing. *RECV* is used to implement the event-driven mechanism of NC. *SEND* is used to set the neuron type, the fired neuron ID, and the 16-bit data transmitted between NCs. *FINDIDX* is an instruction designed to accelerate multi-cycle sparse weight lookups based on bitmaps. *LOCACC* is used to accumulate current. *DIFF* is used to accelerate first-order PDEs of the form $v = \tau v + c$. These instructions are designed to take into account the characteristics of the brain-inspired computing model structure and temporal-spatial double-sparse spike [37], [38], as well as the dynamic behavior based on partial differential equations. In addition, NC also supports a variety of arithmetic logic calculations, conditional operations, comparison operations, etc., in order to realize more complex neuron behaviors and on-chip learning algorithms.

In NC, each neuron has independent parameters such as weights and membrane potential threshold. The random sparse spike events will activate different destination neurons to participate in the calculation, making the corresponding data have a short life cycle and no data temporal locality. locality

TABLE I
INSTRUCTION SET OF TAIBAI

| Instruction | Operands | Examples |
|---|---|---|
| RECV | None | Neuron core hangs up and waits for spike events to arrive |
| SEND | Register | Sending 16-bits value, fired neuron ID and neuron type |
| FINDIDX | Register | A bitmap-based sparse weight accelerated lookup instruction |
| LOCACC | Register, memory | Current accumulation instruction |
| DIFF | Register, memory | Accelerating the solution of first-order partial differential equation |
| ADD/SUB/MUL | Register, immediate | FP16 and INT16 arithmetic instructions for addition, subtraction and multiplication |
| ADDC/SUBC/MULC | Register, immediate | Conditional execution of arithmetic instructions for FP16 and INT16 operations of addition, subtraction and multiplication |
| AND/OR/XOR | Register, immediate | Logical operation instructions for AND, OR and XOR |
| CMP | Register, immediate | FP16 and INT16 comparison instructions |
| MOV | Register, immediate | Data movement instructions |
| LD/ST | Register, immediate, memory | Read/write memory instructions |
| B/BC | Register, immediate | Branch and conditional branch instructions |

Traditional load-store microarchitectures leverage temporal of data for optimization, frequently performing read–compute–writeback operations in SNN calculations. These architectures not only lead to additional power consumption but also increase computational cycles, making them clearly unsuitable for brain-inspired chips. To this end, we design a reg-mem seven-stage pipeline microarchitecture (Fig. 3) to improve the running efficiency and throughput of instructions, and to eliminate the adverse effects of short data life cycle and randomness of sparse spike event. Output events mem is used to store the types of output data, the IDs of fired neurons, and data. The input event buffer caches the neuron IDs, axon IDs, and data corresponding to the event so that NC can complete the model calculation.

The dynamic process of the neuron model is divided into two stages: the integration stage, where dendrites receive and integrate spike signals, and the membrane potential update and spike firing stage. Since the processing of SNN depends on layer-wise timestep synchronization, we decouple the SNN computation at each timestep into two stages (INTEG and FIRE), and simplify the chip control logic through a model pipeline parallel computation mechanism. In the INTEG stage, after the NC receives a spike event, it immediately completes the corresponding current accumulation process. Otherwise, it remains in a resting state, thereby fully leveraging the sparsity of spikes to reduce power consumption overhead. After a period of current accumulation, there are no spike events in the NoC, and the chip enters the FIRE stage. The NC will update the membrane potential according to the neuron dynamics equation, and mark the fired neuron ID in the output event memory. As for on-chip learning algorithms, the NC completes the weight update process during the FIRE stage. In addition, the NC supports both spike and floating-point input and output modes. The floating-point input capability enables the direct processing of non-spiking data on the chip, the floating-point output of can be used to output information such as membrane potential and model errors, thereby enhancing the support capabilities of the model and on-chip learning capabilities.

*C. Router with hybrid-mode routing*

In biological brain models and brain-inspired models, the postsynaptic neurons of a neuron exhibit strong small-world characteristics [39], presenting dense local and sparse global connectivity [40]. To this end, we design a destination-driven router that supports three routing modes: point-to-point, regional multicast, and broadcast. We use a 2D mesh topology as the architecture of the NoC. In terms of routing algorithm design, we use XY routing algorithm for point-to-point transmission and use the tree-based broadcasting algorithm for broadcasting. As for regional multicasting, the router will automatically select the shortest path to the regional boundary based on the current node location, and then use the tree-based multicasting algorithm within the region to minimize the propagation delay and the number of data packets. The 64-bit packet includes information of type, phase, tag, index, destination area, and payload. The type field not only encodes the three spike-packet routing modes discussed above, but also specifies memory-access modes, enabling model configuration and run-time monitoring of model status. The phase field is used to mark the work stage of multicast and broadcast, and the tag and index are used to index the first-level fan-in table of CC.

*D. Scheduler for efficiently supporting topologies and scheduling events*

To efficiently support diverse network topologies while taking core scheduling efficiency into account, we designed the scheduler architecture. Most network topologies can be categorized into three types: sparse connections, full connections, and convolutional connections. By analyzing the commonalities and differences among these topologies, we have devised a multi-level network topology representation scheme, including three mechanisms of *incremental addressing of neurons in the fully connected layer*, *parallel sending mechanism*, and *decoupled convolution weight addressing*. For network topologies that are not explicitly mentioned, such as recurrent connections, we can equivalently convert these topologies into existing ones, thereby enhancing the adaptability to diverse network topologies. Considering that skip connection breaks the layer-wise timestep synchronization of the chip, we propose a resource-friendly skip connection representation scheme through hardware-software co-design.

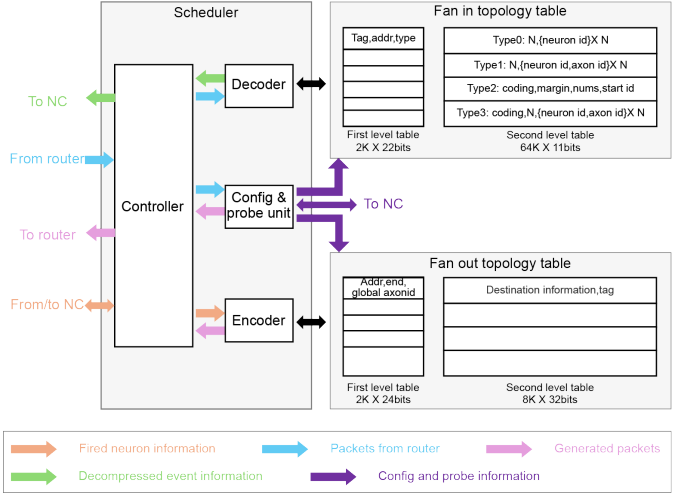

Fig. 4. The block diagram of scheduler.

1) *Scheduler:* The scheduler in CC is primarily used to schedule spike events and NCs, thereby achieving efficient parallel computation under sparse event conditions. When the scheduler receives a spike event packet, it determines the corresponding axon ID and target neuron ID through the fan-in topology table, and initiates the corresponding NC for computation. When an accessing-memory data packet is received, the scheduler completes the initial configuration of the model based on the packet information. Alternatively, it reads the neuron information stored in the NC and generates a data packet to return to the host machine. When a neuron needs to fire a spike, the scheduler queries the fan-out topology table, generates a spike event packet, and sends it to the router. The specific architecture of the scheduler is shown in Fig. 4.

2) *Topology on hardware:* The network topology is divided into fan-in and fan-out from the perspective of the connection between neuron inputs and outputs, and they are stored in the fan-in topology table and fan-out topology table, respectively. Both tables (Fig. 4) are organized as 2-level tables with sparse connectivity representation and indexing capabilities, thereby saving on-chip storage resources required for topology representation. We define the first-level table as the Directory Table (DT), where each element is a Directory Entry (DE) used to index the second table. The second-level table is the Information Table (IT), where each element is an Information Entry (IE).

The DE and IE of fan-out table also include the global axon ID and routing information respectively, both used to generate data packets. We use fired neuron ID to address the DE of fan-out table. Since the regional multicast algorithm uses rectangles for regionalization, there will be non-targeted CCs within the region. For this reason, fan-in DE includes an identifier tag so that the scheduler can determine whether to continue processing the packet based on the tag of the packet. The fan-in IE is the information related to the target neuron and target axon, and the design of this part determines the efficiency of network topology coding. We analyze the commonalities and differences in connection modes of mainstream neural network topologies (convolution, pooling, full connection, recursive connection, sparse connection), and design four types of fan-in IE to improve the efficiency of topology encoding and efficiently support various network topologies.

2) *Sparse connection:* For sparse connection, two types of fan-in IE are provided to meet different requirements for sparse weight decoding speed and topology representation resource overhead. **Type 0**: IE is the IDs of the neurons that are connected to the upstream neuron, and NC decodes the sparse weight address through the global axon ID (Fig. 5(a)). This type of table is mainly used for scenarios with storage resource constraints. The table only stores neuron ID information and can be used as a basic method for describing neuron connections. For example, the pooling topology uses this type of table. In addition, this scheme can also be used for sparse connection with low sparsity. Sparse weights are compressed based on the bitmap, and the global axon ID is used to look up weights by using *DINDIDX* instruction. **Type 1**: IE includes multiple sets of neuron ID and local axon ID (Fig. 5(b)), and NC directly reads the weights through the local axon ID and completes the accumulation of the currents of the relevant neurons. This type of table is mainly used in high-throughput scenarios. It does not need to obtain weights through sparse weight decoding operations with high latency, which improves computing efficiency.

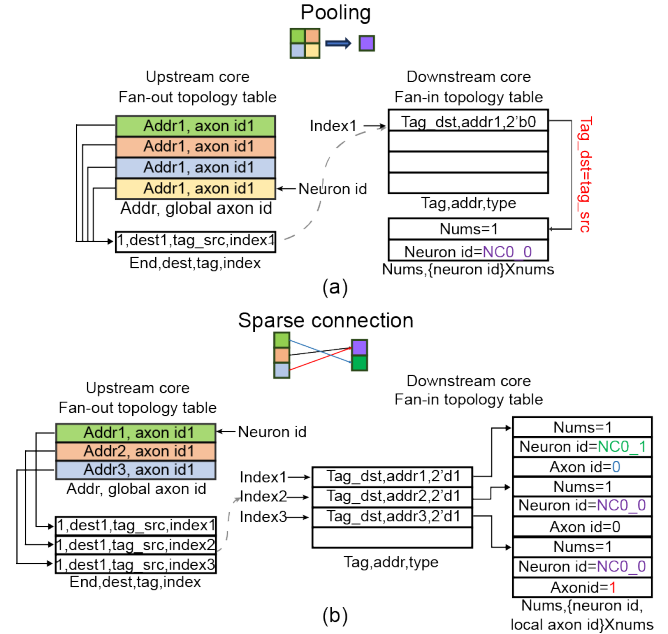

Fig. 5. Sparsely connected representation scheme, and the example of (a) pooling and (b) sparse connection.

3) *Full connection:* For full connection, each upstream neuron has the same destination neuron, and the weight address of the destination neuron is only related to the upstream neuron ID. In this connection pattern, a spike event will cause the computation of all destination neurons. We proposed a **type 2** IE, on the one hand, the mechanism of *incremental addressing of neurons* only requires four entries to represent all destination neuron (coding, margin, number of accumulations, and starting neuron ID, as shown in Fig. 6). On the other hand, the *parallel sending* mechanism evenly distribute downstream neurons to multiple NCs and send events to different NCs in parallel according to the coding mask to enhance computational parallelism.

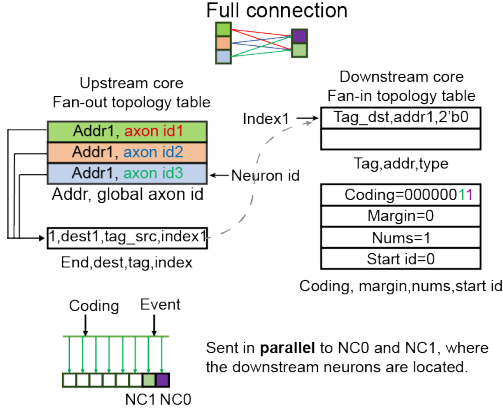

Fig. 6. Topology representation scheme with parallel sending and incremental addressing mechanisms for full connection.

*5) Convolutional connection:* For convolutional connection, upstream neurons at the same position in different channels have the same destination neurons, and the destination neurons of the same channel share a set of filters. However, decoding the shared filter weights corresponding to the events requires a complex algorithm with high computational overhead [41]. To this end, we *decouple the convolutional weight addressing process* and design **type 3** IE (Fig. 7), including mask, numbers, neuron ID, and local axon ID. The global axon ID of fan-out DE is the channel ID of the upstream neuron, and local axon ID is the address of the filter weights. A simple polynomial shown in (4) can be used to complete the decoding of the weight address based on the global-local axon ID. This design makes the number of IE related to the number of single-channel neurons, rather than the number of channels, which improves the degree of sharing in the topological representation of multi-channel feature maps. Similar to the **type 2** IE, this IE also supports the parallel transmission of spike events to multiple NCs and the parallel computation of neurons with multiple channels.

$$w_{addr} = axonid_{global} * k^2 + axonid_{local} \quad (4)$$

6) *Skip connections:* Skip connections disrupt the layer-wise timestep synchronization of the brain-inspired chip pipeline, resulting in network layers needing to receive multiple sets of data processed at different timesteps. The traditional method adds a set of relay neurons to cache spikes for synchronization, but this will bring a very large resource consumption (Fig. 8(a), (b)). Alternatively, a set of fan-out DE is used only to represent skip connection information, which results in idle resources when deploying other connections. To this end, when supporting skip connections, we reuse the neuron type of NC in output event mem as spikes that need to be delayed fired (Fig. 8(c)). The delayed and non-delayed spikes share the fan-out DT, and only differ in direction when addressing the fan-out IT, in order to make full use of the fan-out table. In the model design, the corresponding spikes are cached in NC according to the number of layers spanned by the skip connection, and delayed spikes are fired at the correct time for synchronization. In addition, we fuse the operation of downsample to the destination NC (core4 in Fig. 8(b)) to further reduce the core requirements.

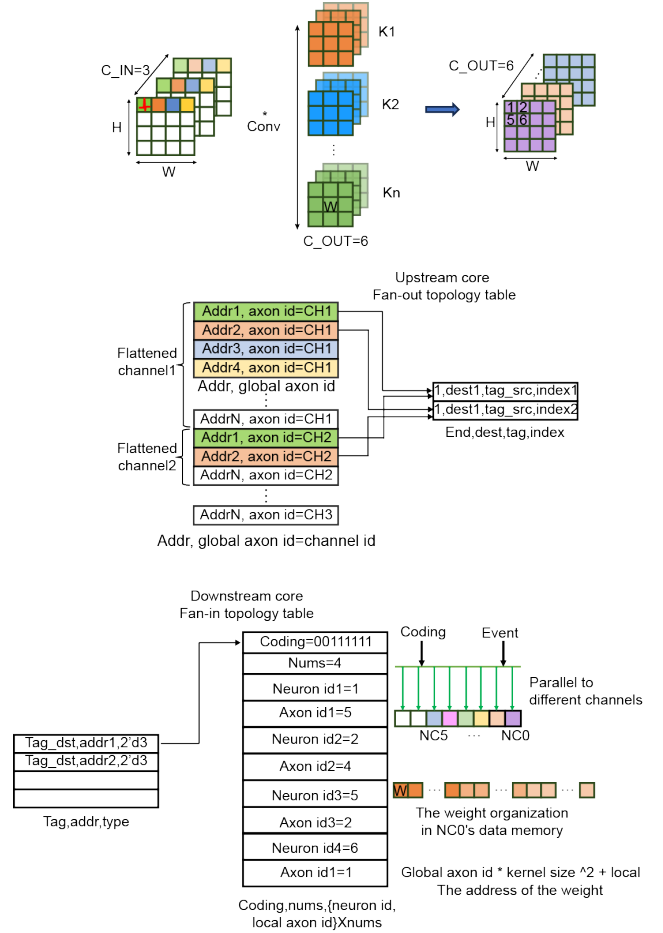

Fig. 7. Topology representation scheme with parallel sending and decoupled convolution weight addressing mechanisms for convolutional connection.

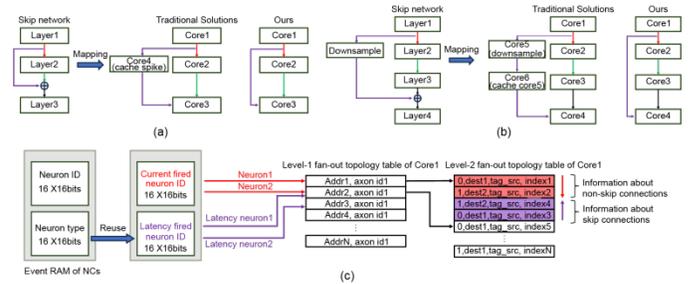

Fig. 8. The representation scheme for skip connection.

## IV. COMPILATION AND MAPPING FRAMEWORK

This section introduces the framework that carries brain-inspired models from high-level description to efficient execution on TaiBai. Section IV-A traces the life-cycle of a spike event inside the chip, illustrating how packets are decoded, routed, and processed by neuron cores. Section IV-B then explains how the architecture accommodates new neuron dynamics, learning rules, and the scale of neuron fan-in and fan-out. Finally, Section IV-C describes the end-to-end software flow that parses user models, partitions and schedules the locations, and emits binaries for the instruction set. Together, these stages establish a seamless path from algorithm to hardware while preserving programmability, extensibility and performance.

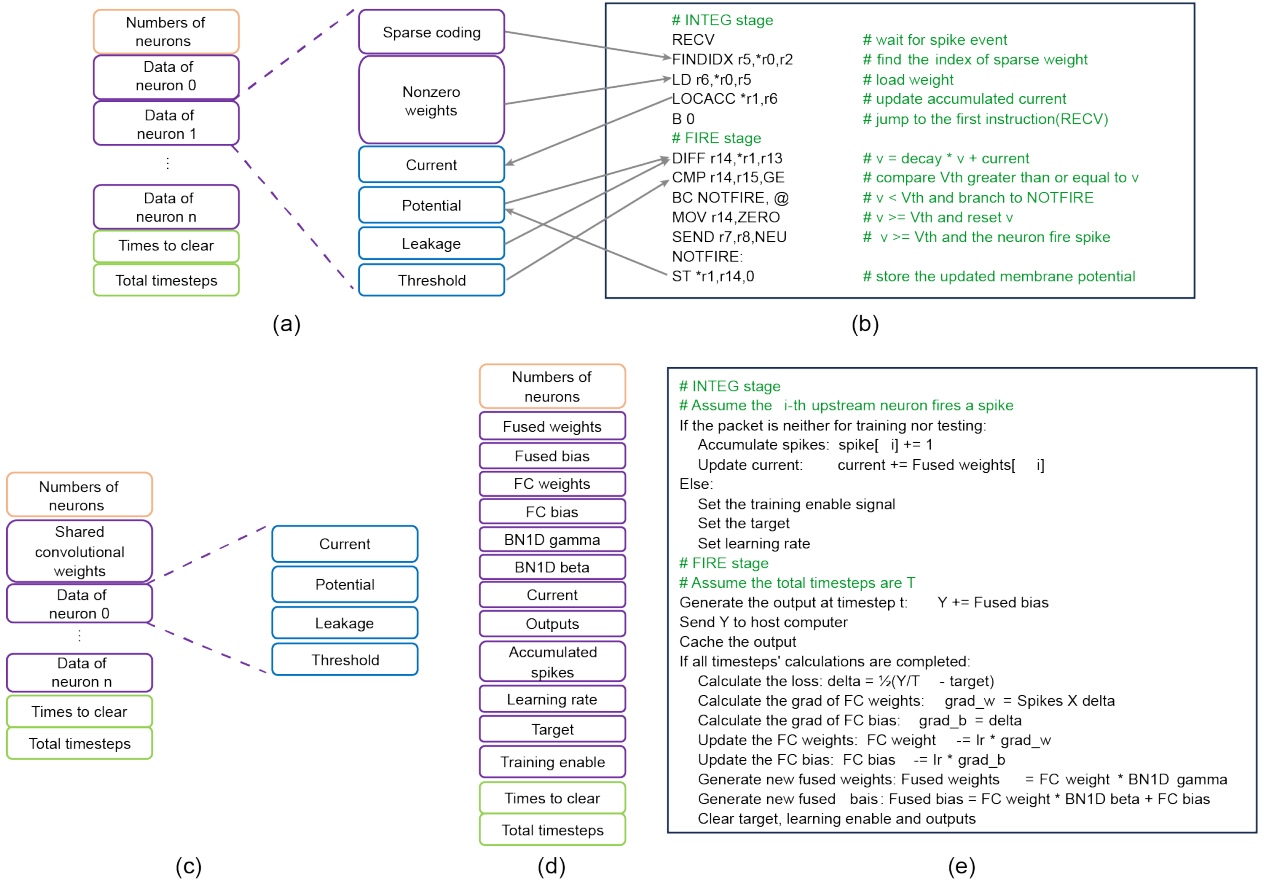

Fig. 9. Data structure and model representation in TaiBai. The (a) structure and (b) calculation code of LIF neuron model with sparse connections are the basic models of TaiBai, which can evolve (c) convolutional models with shared weights, (d) and (e) on-chip learning algorithms, etc.

*A. On-Chip Runtime Workflow*

The TaiBai chip has three working stages in total: initialization (INIT), INTEG, and FIRE. As shown in Fig. 10, when in the INIT stage, all NCs are in resting state. The scheduler reads the accessing-memory data packet, and then writes the model and topology into the corresponding memory. All neurons in the network layer are mapped to specific NCs and will run in parallel in the form of a pipeline on the chip. Each NC processes the data and mapped neurons corresponding to the timestep, and sends the results to the destination core according to the model topology at the same time. After completing initialization, the core enters the SNN computation process and repeatedly executes the INTEG-FIRE stages, sequentially completing current accumulation and model updates. One complete INTEG-FIRE stage is one timestep in SNN. The TaiBai compiler automatically selects the number of cycles per timestep based on the complexity of the model. This ensures correct model computation while balancing computational latency and used cores. In order to facilitate the observation of the model running status, TaiBai allows the sending of accessing-memory data packets in the FIRE stage, and the scheduler sends the required information to the host computer.

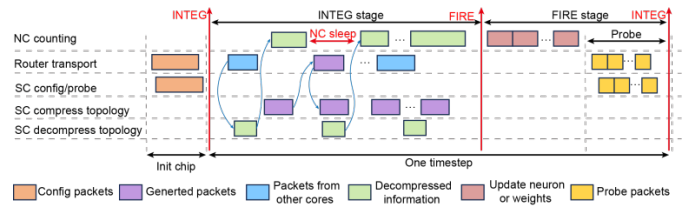

Fig. 10. Workflow of the chip.

*B. Model Extensibility*

We build data structure and model representation in TaiBai for neuron models and learning algorithms extensibility. Fig. 9(a) and (b) show the data structure of sparsely connected LIF neurons and the model calculation code, respectively. The arrow in the middle indicates the data dependency relationship corresponding to the code. Each neuron has its own data area, including compressed sparse weights and neuron variables. When an event wakes up the NC calculation, the model calculation is quickly completed using a small number of high-performance instructions (5 instructions in INTEG stage and 7 in FIRE stage) then the NC waits for the next event to arrive. LIF neurons are capable of sharing convolutional weights shown in Fig. 9(c).

For on-chip leanring, Fig. 9(d) and (e) show the data structure and pseudocode of the learning algorithm for the BCI cross-day decoding task, respectively. We fuse BN1D and FC into one FC operation to reduce the amount of computation, and the corresponding parameters are expressed as fused weights and fused bias. Since the backpropagation through time algorithm requires spikes of each timestep to adjust the weight, directly storing the spikes will introduce huge storage overhead, while storing compressed spikes (e.g. based on bitmap) will introduce additional decompression time. We have optimized the on-chip learning algorithm by balancing storage overhead and computational speed. The values of multiple-timestep spikes are accumulated during forward propagation to reducing the storage requirements. During the backward propagation process, the accumulated spikes are used instead of timestep-by-timestep spikes, which speeds up on-chip learning.

Although TaiBai efficiently supports various network topologies, the resources for storing topological information are limited (TaiBai constrains each neuron to have a maximum of 2K fan-ins). This brings challenges when deploying models with larger fan-in and fan-out. We enhance the connectivity capabilities of TaiBai by designing fan-in and fan-out expansion methods. When performing fan-in expansion, multiple partial sum (PSUM) neurons need to be deployed, and these neurons compute a part of the accumulated current. When the membrane potential is updated, the local accumulated current is transmitted to the spiking neuron, which performs the updating of the membrane potential and the firing of spikes. Since most existing architectures do not support the transmission of neuron data within the NC, the two types of neurons must be placed in different cores. The PSUM neuron only receives spike events and the spiking neuron only receives accumulated current. This architecture increases the number of cores required and the computational latency when expanding fan-in. Since TaiBai supports data transmission within NC, the spiking neurons can receive not only accumulated current but also spike events, reducing resource overhead and computational latency (Fig. 11).

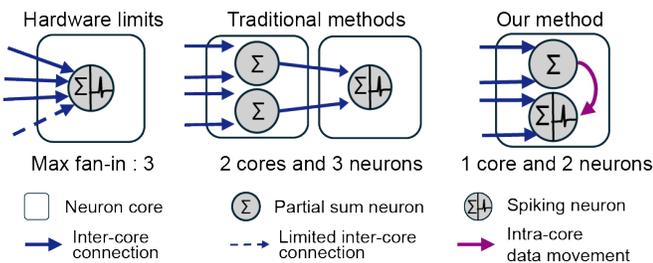

Fig. 11. Fan-in capability expansion.

In fan-out expansion, the original neuron needs to be decomposed into multiple neurons with local destination areas. These neurons have the same firing time, and the sum of all destination areas is the original destination area, which can be deployed intra-NC or inter-NC. The intra-NC expansion scheme reduces the number of configurable neurons and increases the fan-out information of the previous layer. The inter-NC expansion scheme increases the computational latency. It is necessary to select the appropriate fan-out expansion plan according to the requirements. In terms of chip-scale expansion, data proxy units around the chip enable intra-chip and inter-chip packet conversion, and use the same routing algorithm as the intra-chip routing for inter-chip data routing and forwarding. These various expansion methods make TaiBai easy to deploy larger scale models.

*C. Compiler stack*

TaiBai is dedicated to enhancing the programming flexibility required for SNNs, optimizing the architecture design for layer-wise and channel-wise network partitioning. As models become more complex and scale continues to expand, deploying models across thousands of cores presents a significant challenge. We have designed a compiler stack that takes into account both programming flexibility and hardware execution efficiency, enabling optimized deployment of brain-inspired computing tasks on the TaiBai chip to make full use of its performance. In addition, it also has a behavioral-level chip simulator, which can be used as a brain-inspired algorithm verification platform, supporting offline simulation of the algorithms on the chip and statistical runtime information. The simulator can quickly evaluate the model's operating power consumption, throughput, and resource usage on the chip. It also simulates the operation process of each unit, providing references for hardware debugging.

Deploying SNN models to multi-core brain-inspired chips mainly includes two steps: network partition and core placement. Network partition assigns neurons to each core, and core placement will affect communication congestion and computational latency. For the SNN models that need to be deployed, we can accept various front-end representations (Fig. 12(a)). The compiler stack first extracts the basic operators of the model and fuses multiple operations of a layer into one operator, such as fusing convolution and BN or pooling into convolution (Fig. 12(b)). Then, neurons on each layer are assigned to specific cores in channel order (Fig. 12(c)), and the initial placement of the cores is completed based on the zigzag curve. Each core contains a computational model, model parameters, synaptic parameters, network connection topology information, and the region to which the core belongs. The resource optimizer can merge cores with the same operators at different layers, solving the problem of low utilization of some core resources when performing network partitioning by channel, thus reducing the number of cores required for the model. For core placement, we obtain the number of data packets sent by each core through the chip simulator. Using task throughput and resource utilization as optimization objectives, we employ greedy algorithms to optimize core placement (Fig. 12(d)). Finally, the code generator will generate the assembly code of the computational model, and generate configuration data packets of the computational model, model parameters, and model topology. The packets will be sent by the host computer to the hardware platform to complete the chip configuration (Fig. 12(e)).

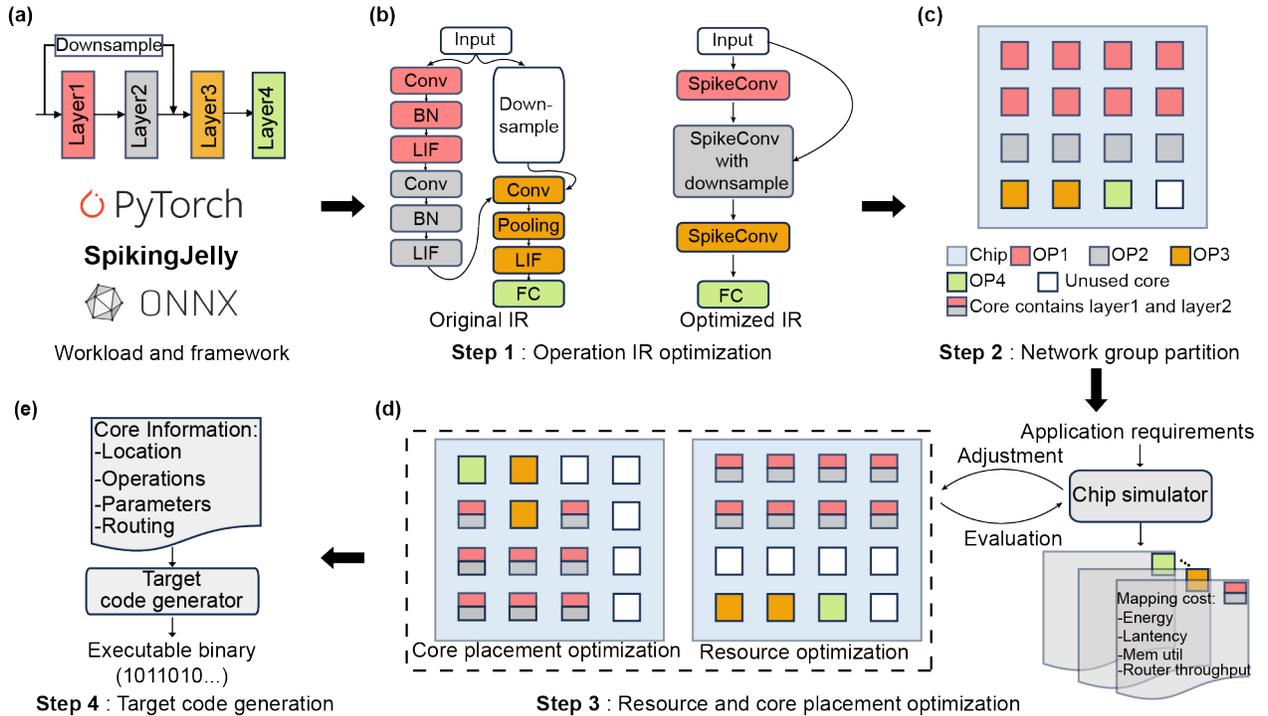

Fig. 12. Compiler stack. The process of deploying (a) a workload built on multiple frameworks with multiple network topologies to TaiBai consists of four steps. Step 1: (b) Obtain the operation IR of the workload, optimize it, and generate the operation IR suitable for TaiBai. Step 2: (c) Partition the neurons corresponding to the optimized IR into different cores. Step 3: (d) Based on the application's requirements (used cores, throughput, etc.), merge cores from different layers to reduce the number of cores, optimize core placement to enhance throughput, and evaluate the process using the chip simulator to guide the optimization. Step 4: (e) Generate a chip-recognizable binary file based on the results of Step 3.

## V. EXPERIMENTS

### A. Chip Evaluation

We developed Verilog code to implement the architecture. A chip with 11*12 CCs prototype FPGA verification platform was built using six Xilinx Virtex UltraScale+ VU13P boards and two Xilinx MPSOC ZU19EG boards. One VU13P can simulate 5*8 CC arrays. The host computer (ThinkPad neo14, Lenovo) and VU13P communicate through ZU19EG. The FPGA boards communicate via optical modules. When performing the application demonstration, the host computer sends the model and sample data to ZU19EG via Ethernet. ZU19EG will continuously send sample data to VU13P according to the configured INTEG and FIRE intervals, and collect the operation results of VU13P. The host computer then analyzes the results and repeats the above process. It should be noted that in the applications, we only need one VU13P borad (capable of simulating 40 CCs) to complete these challenging tasks due to the efficient resource optimization strategy and topology representation scheme. In addition, in order to compare with other chips, TaiBai is synthesized based on SMIC 28-nm low voltage threshold CMOS process and operates at 500MHz.

### B. Experiment Setup

*1) Implementation of the compiler stack:* We developed an assembler for the TaiBai instruction set based on flex and bison, and developed the mapper using Python for optimizing operator intermediate representation, network partition, core placement optimization, and resource optimization. Genetic algorithms or simulated annealing algorithms are used to optimize core placement to reduce congestion, and multi-network fusion strategies are used to reduce core resource usage. In addition, we developed a behavior-level chip simulator using Python, which can directly obtain information about the resource utilization, running rate, router throughput, operating power consumption, and accuracy of the SNN models on TaiBai.

*2) Benchmarks:* We use the chip simulator to obtain the running power consumption and running time of TaiBai. For GPU power consumption, we record the static power during an extended period with no workload. Then we record the power while the model is running and calculate the average power. All power consumption is obtained by the pynvml API provided by NVIDIA. Execution time is obtained by running the model 15 times on the GPU and calculating the average.

In benchmarks compared with the GPU, we used three common SNN networks with network structures shown in Table II. We employed maximum throughput mapping as optimization objectives and compared accuracy, power consumption, and energy efficiency. In addition, we analyze the impact of different mapping optimization strategies on the number of used cores and energy efficiency. These models all use the LIF neuron model (all network models use the LIF model if not specified). In the analysis of the network topology representation, we use the standard ResNet [42] and VGG [43] network architectures, replacing their activation functions with spiking neurons.

TABLE II
NETWORKS OF SNN FOR TAIBAI EVALUATION

| Name | Network structure | Input shape |
|---|---|---|
| PLIF-Net | Input-256c3p1X3-mp2-256c3p1X3-mp2-fc4096-fc10 | 32*32*3 |
| 5Blocks-Net | Input-mp2-16c3-[16c3p1X2]-mp2-[16c3p1X2]-mp2-[16c3p1X2]-mp2-[16c3p1X2]-mp2-[16c3p1X2]-mp2-fc11 | 128*128*2 |
| ResNet19 | Input-64c3-[128c3p1X2]X3-[256c3p1X2]X3-[512c3p1X2]X2-fc256-fc10 | 32*32*3 |

*3) Algorithms and datasets for multiple applications:* We demonstrated three complex tasks on the chip prototype system, namely ECG signal recognition, speech recognition and BCI decoding. The models of these applications cover a variety of neuron models, network topologies and on-chip learning algorithms.

For datasets, the ECG signal recognition task uses the QTDB [44] dataset, which consists of 759 sequential data, each consisting of six different characteristic waveforms: P, PQ, QR, RS, ST, and TP. We use level-crossing coding to convert the continuous values of each channel into two independent positive and negative spike sequences. The converted data dimension is 4*1301, where 4 is the number of input channels and 1301 is the timestep. For the speech recognition task, we use the SHD [45] dataset, which includes 10,000 recordings in English and German (digits 0 to 9). We employ the method from the original paper to sample the spike sequences $T$ times at a time interval of $dt$, where $T$ represents the timesteps. Each recording is converted into a 700×$T$ binary matrix. For the BCI data, we used macaques as experimental subjects to perform four types of hand movements for visual tasks, while simultaneously recording 128-channel neural signals from the motor cortex (M1) at a sampling rate of 30 kHz. We collected data over 8 days, conducting an average of 300 experiments per day. Each recorded experimental dataset was subjected to Butterworth filtering and a 20 millisecond time window, resulting in data with dimensions of 128×50. The 8 days of data were evenly divided into two groups, with one day's data used for training in each group and the remaining three days used for cross-day decoding experiments.

For ECG signal recognition, we use the SRNN model proposed by Yin [19], which features a recurrently connected hidden layer and an output layer. The hidden layer employs the ALIF neuron model, which features an adaptive threshold that increases after each emitted spike and then decays exponentially. The output layer employs the LIF neuron model, which determines the band of the signal in real time by assessing the membrane potential of output neurons at each timestep. In addition, we deployed a model with the same network structure but exclusively using LIF neurons.

For speech recognition, we use the DHSNN with the dendrite model proposed by Deng [15]. The network consists of a hidden layer and an output layer, with 64 and 20 neurons respectively. The hidden layer uses the DH-LIF neuron model with 4 dendrite branches. The output layer employs a variant of the LIF neuron, which does not exhibit spike firing and membrane potential resetting behavior. During the deployment of this model, a single neuron with four dendrites requires a total of 2,800 fan-ins (hardware limited to 2,048). To address this, we adopted a fan-in expansion approach, calculating the currents of the four dendrites separately within a single core. Subsequently, a spiking neuron located in the same core integrates the dendritic currents and completes spike firing. Additionally, we deployed a model with the same network structure but without dendrites.

For BCI decoding, we designed an SNN with 16 sub-path networks to speed up parallel processing. Each sub-path consists of three modules: a linear transformation module for initial signal processing, a channel attention module to enhance cross-channel features, and a temporal convolution module to extract dynamic temporal features from signals. The outputs of the three modules are fused using the Hadamard product and matrix addition. The outputs from multiple sub-path networks are concatenated together and then passed through LIF, 1D batch normalization, and a fully connected layer to obtain the model output. The decoding result is determined by evaluating the values of the output layer. During cross-day decoding fine-tuning on TaiBai, only 32 samples from the training set are used to adjust the weights of the fully connected layer, employing backpropagation as the learning algorithm, and the performance of the fine-tuned model is validated on the test set. This model has a relatively complex network topology and computational modules so we employed various optimization strategies to deploy this model, ultimately reducing core resources by 3.4 times.

TABLE III
CHARACTERISTICS AND PARAMETERS OF TAIBAI

| Features | TaiBai | Features | TaiBai |
|---|---|---|---|
| Technology | 28nm | Neurons | 264K |
| Clock | 500MHz | Synapses | 6.95M $\sim$ 297M |
| Chip area | 248mm$^2$ | Intra-chip performance | 363MSE/S |
| Power | 1.83W | Inter-chip performance | 322GSE/S |
| Supply voltage | 0.9V | Neuron model | Fully programmable |
| Bit width | 16 | On-chip learning algorithms | Fully programmable |

*C. Experiment Result*

*1) Performance of the TaiBai chip:* Table III lists the basic characteristics and parameters. The chip can support a total of 264K neurons and 6.95M (sparse mode) $\sim$ 297M (convolutional multiplexing mode) synaptic weights. The physical layout of the chip and each CC is shown in Fig. 13(a) and Fig. 13(b). The power breakdown of TaiBai is shown in Fig. 13(c). During operation, the memory module (including the accessing memory process of the NCs and schedulers) consumes the most power (70.3%). We use spike events per second (SE/S) to evaluate the communication capability of the chip. Through the proxy unit and the inter-chip high-speed interface, the chip can achieve an inter-chip communication bandwidth of 363MSE/S and an intra-chip bandwidth of 322GSE/S. Typically, it can achieve a peak performance of about 528 giga synaptic operations per second (GSOPS) with a power consumption of 1.83 W.

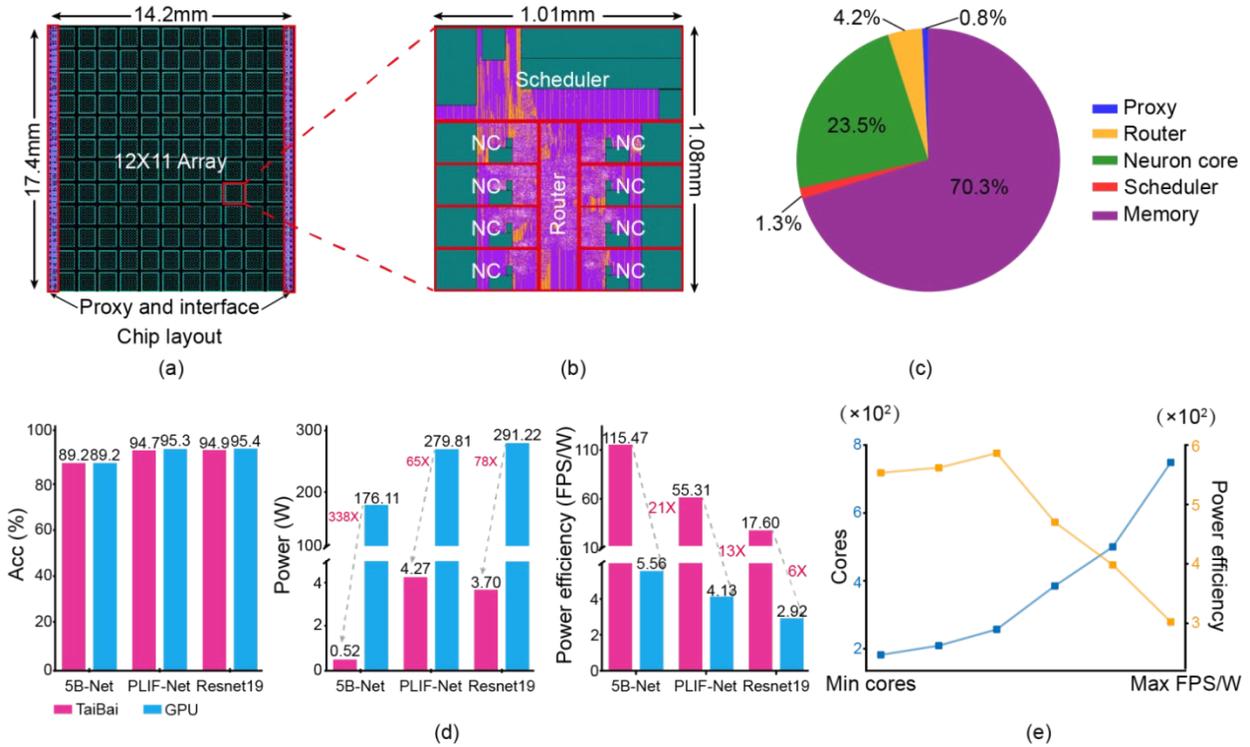

Fig. 13. Chip evaluation and performance. (a) Chip layout. (b) Layout of CC and router. (c) Power breakdown of TaiBai. (d) Benchmarking three representative SNN models on TaiBai (pink) versus a GPU (blue). (e) Compiler controlled mapping of a SNN on TaiBai illustrating the performance versus power trade off.

TaiBai can support various network topologies. To fully evaluate the performance of TaiBai, we deployed several common network models. Based on the chip simulator, we compared the accuracy, power consumption, and power efficiency with NVIDIA GeForce RTX 3090 in Fig. 13(d). In three standard neural network benchmark tests, TaiBai can achieve an accuracy similar to that of the GPU, with the power consumption reduced by 65 to 338 times and the power efficiency improved by 6 to 20 times. We observed that TaiBai's power efficiency varies significantly across different models compared to GPU performance. On the one hand, the PLF-NET and ResNet19 models have a large number of neurons, requiring dozens of chips to run the models. The massive number of intra-chip and inter-chip packets reduces the model's throughput. On the other hand, the spike firing rate of the latter two models is higher than that of the first model (13% vs 8%), resulting in an increase in the dynamic power consumption of the chip. While the spike firing rate has little to no impact on the power consumption of GPUs, which are based on dense tensor computations. Therefore, the power efficiency of the latter two models decreased.

Furthermore, we also deployed a same SNN with different task requirements for comparison. For resource-aware scenarios, minimizing the number of cores was used as the optimization goal. For throughput-aware scenarios, maximizing throughput was the optimization goal. As shown in Fig. 13(e), the horizontal axis indicates from minimizing the number of cores to maximizing the throughput. The blue line indicates the number of cores used, and the yellow line represents energy efficiency. As the demand for throughput increases, the number of cores increases by 4 times (182 to 749), and energy efficiency decreases by 1.7 times (6190 to 3590). In practical applications, we can trade off the resources, fps, and power according to scenario requirements.

*2) Comparison with the state-of-the-art brain-inspired chips:* As shown in Table IV, we divide brain-inspired chips into two categories based on programmability: those on the left support specific models, and those on the right have different degrees of programmability. Programmable means that the chip is designed with special data paths and instructions for certain models. Although other models can be implemented through instruction combination, the types of models supported by this method are limited. For example, the on-chip learning rules of Loihi2 are only applied to pre-, post-, and generalized "third-factor" traces. Fully programmable is defined as the ability to implement any models, and both SpiNNaker and TaiBai belong to this category.

TABLE IV
COMPARISON OF TAIBAI WITH STATE-OF-THE-ART BRAIN-INSPIRED CHIPS

| Processor | TrueNorth [4] | Loihi [6] | Tianjic [7] | PAICORE [26] | SpiNNaker [5] | Loihi2 [22] | Darwin3 [27] | TaiBai (This work) |
|---|---|---|---|---|---|---|---|---|
| Technology(nm) | 28 | 14 | 28 | 28 | 130 | 7 | 22 | 28 |
| Die area(mm$^2$) | 430 | 60 | 14.44 | 537.98 | 102 | 31 | 358.527 | 248 |
| # of cores | 4096 | 128 | 156 | 1024 | 18 | 128 | 575 | 1056[a] |
| # of neurons | 1M | 128K | 39K | 1.83M | - | 1M | 2.25M | 264K |
| # of synapses | 256M | 128M | 9.75M | 4.45G | - | 123M | - | 6.95M $\sim$ 297M |
| Precision | 1-bit | 1-to-9-bit | 8-bit | 1-bit | 32-bit | 1-to-9-bit | 1/2/4/8/16-bit | 16-bit |
| Multicast delivery | No | Yes | Yes | Yes | Yes | Yes | No | Yes |
| Frequency | Asynchronous | Asynchronous | 300MHz | 24M-600MHz | 180MHz | Asynchronous | 333MHz | 500MHz |
| Neuron models | LIF | LIF | LIF | LIF | Fully programmable | Fully programmable | Programmable | Fully programmable |
| Synapse models | CUBA Delta | CUBA Delta | - | - | Fully programmable | - | Programmable | Fully programmable |
| On-chip learning | No | STDP | No | STDP | Fully programmable | Programmable | Programmable | Fully programmable |
| Power(W) | 0.042-0.323 | - | 0.95 | 0.01-9.97 | 1 | - | - | 1.83 |
| Energy per SOP | 26pJ | 23.6pJ | 1.54pJ | 0.19pJ | 11nJ | 7.8pJ | 5.47pJ | 2.61pJ |

[a] TaiBai contains 132 CCs and each CC contains 8 NCs.

Although the number of neurons is an important specification parameter of brain-inspired chips, we found that due to a lack of or low programmability, the significance of directly comparing the number of neurons is diminished. For example, TrueNorth has to use multiple neurons to simulate the behavior of complex neurons. Similar comparison is also reflected in the number of synapses. Some existing designs calculate the number of synapses by accounting for reused convolutional weights, resulting in an extremely large count. Energy per SOP reflects the chip's energy efficiency. TaiBai exhibits excellent performance among programmable brain-inspired chips with an energy consumption of 2.61 pJ. PAICORE achieves an impressive energy efficiency of 0.19 pJ. However, this is attained by sacrificing data bitwidth (1-bit). In contrast, TaiBai utilizes both FP16 and INT16 data formats, offering higher computational precision. In terms of chip applications, to the best of our knowledge, TaiBai is the first brain-inspired chip capable of using multiple heterogeneous models to perform tasks.

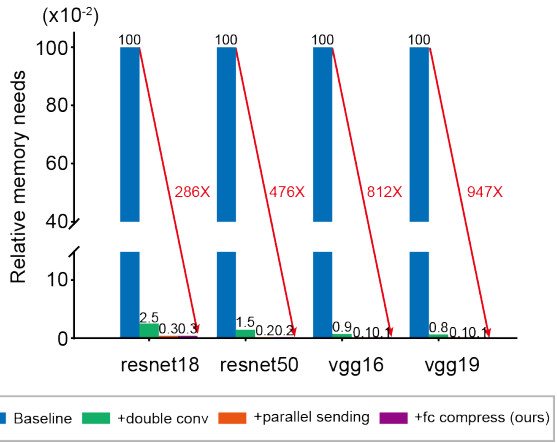

Fig. 14. Efficiency of network topology representation on conventional models.

*4) Efficiency of the network topology representation:* The network topology representation scheme we proposed has three main contributions. Incremental addressing of neurons in the fully connected layer: All neurons are addressed by only 4 entries, regardless of the number of neurons in the fully connected layer. Parallel sending mechanism: This mechanism allows the same event to be sent to $N$ NCs in parallel. Without this mechanism, the resource requirements of the fan-in table need to be increased by $N$ times. Decoupled convolution weight addressing: In the framework of event-driven computing, based on the shared characteristics of convolution operations, the convolution weight addressing process is converted into two parts: global axon ID and local axon ID. This method can quickly calculate the weight corresponding to the spike event through a polynomial and save the storage requirements of the axon ID in the fan-in table. If decoupling is not performed, the postsynaptic neuron and the corresponding axon ID corresponding to each upstream spike event need to be stored, and the resource requirements are increased by $C$ times ($C$ is the number of channels of the previous feature maps), which is no different from expanding the convolution to a full connection.

We implemented some benchmark models and get the resource requirements of fan-out table. As show in Fig. 14, the leftmost column of each model represents the fully connected unfolded mode (baseline). Starting from the second column, it represents the addition of the following methods on the basis of the previous column: decoupled convolution weight addressing, parallel sending mechanism, and incremental addressing of neurons in the fully connected layer. The rightmost column is our method. The results show that our method reduces storage requirements by 286 to 947 times compared to the baseline. The method of using incremental addressing of neurons in the fully connected layer in the figure does not lead to a significant reduction in resources, mainly because the connections in the fully connected layer are far less than those in other layers. In addition, we can directly support residual structures and avoid introducing additional resources overhead by duplicating cores. Taking ResNet18 as an example, the number of cores used by our method is 70.3% of that of the duplicating cores method.

*4) Applications for TaiBai:* To demonstrate the multi-scale flexible programmability and on-chip learning capabilities of TaiBai, we selected three typical SNN applications. These applications feature different input data types, network topologies, neuron dynamic, and synapse dynamic models. We implemented these challenging tasks on an FPGA-based prototype platform for the TaiBai chip.

The above models all use minimizing the number of cores as the optimization goal. Fig. 15(a) is the accuracy of the three models on the TaiBai and GPU platforms (where the accuracy of BCI is the average of the multi-day decoding accuracies). From the results, we can see that TaiBai can achieve learning and inference capabilities comparable to GPU. We use the simulator to compare power consumption and power efficiency with the GPU for these three tasks. As shown in Fig. 15(b), the average power consumption of TaiBai is about 0.34W, about 200 times lower than the GPU, and its power efficiency (Fig. 15(c)) is 296 to 855 times that of the GPU. The performance advantage of TaiBai in BCI tasks comes from parallel computing of multiple branch networks, and it takes advantage of sparsity in speech recognition (input spike rate is 1.2%, hidden layer neuron spike firing rate is 2.5%). Although the spike firing rate in the ECG recognition task is high (33%), the fan-out table that allows neurons to compute in parallel improves the task throughput. The results demonstrate that TaiBai not only provides flexible and efficient support for dendritic models, neuron models, on-chip learning algorithms, and various network architectures but also exhibits excellent high energy efficiency.

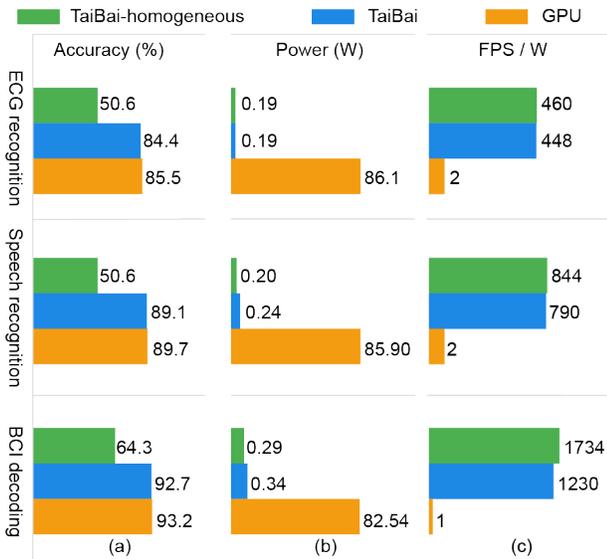

Fig. 15. Applications performance of (a) accuracy, (b) power, and (c) energy efficiency (FPS/W) comparison between TaiBai and GPU. Homogeneous is used to characterize the complexity of the running application.

Additionally, Fig. 15 incudles a *TaiBai-homogeneous* baseline, obtained by deploying SRNN without heterogeneous neurons, DHSFNN without dendrites, and BCI cross-day decoding without on-chip learning. As can be seen from the figure, the heterogeneous neuron model makes it easier to identify the band type of ECG data, the DHSFNN with a dendritic model significantly improves the network performance, and the on-chip learning capability improves the decoding accuracy across days.

## VI. CONCLUSION

In conclusion, we propose a fully programmable many-core brain-inspired chip for various brain-inspired models. In terms of hardware architecture design, we proposed a brain-inspired instruction set and a tightly coupled pipeline structure for the neural dynamic model, low data lifecycle, and data dependency expressed by PDE. A topology representation based on two-level tables is designed for diverse connections of neurons. We design efficient coding schemes for common topologies such as convolution, full connection, and skip connection. An event-driven operation mechanism is designed to address the spatiotemporal sparsity of brain-inspired computing. The unified and efficient routing strategies facilitate large-scale direct expansion of chips. To fully utilize the advantages of the TaiBai chip, we have designed a complete compiler stack and development environment that can deploy various brain-inspired models for different application scenarios based on model characteristics. It greatly simplifies the process from model design to hardware execution, improves the efficiency of system operation and reduces the difficulty of development. The compiler stack completes task compilation end-to-end and supports operator optimization, network partition, resource optimization, core placement optimization, and hardware behavior simulation. In addition, to address some hardware limitations, we provide low-resource and low-latency overhead expansion methods to deploy large-scale models directly.

In general, the TaiBai chip takes into account model flexibility, topology diversity, and scalability of large-scale networks in hardware design, thereby supporting the deployment of various brain-inspired models. It can meet the rapid iterative development and evolution of brain-inspired research. The complete compiler stack ensures the efficient execution of the chip, so as to give full play to the performance advantages of brain-inspired models and chips. Through multiple application demonstrations, we verified the potential and advantages of TaiBai in ECG signal recognition with heterogeneous neurons, speech recognition with dendritic models, and cross-day decoding of BCI with on-chip learning. Our key goal is to further expand the support for more models, covering more diverse network topologies and more complex neural dynamics models. Comprehensively improve the adaptability and expression capabilities of the chip in different brain-inspired computing tasks, thereby providing a platform foundation for a wider range of brain-inspired computing needs. Through the continuous synergistic optimization of software and hardware, we believe that TaiBai will be able to support more and more complex computational models, support the rapid development of brain-inspired research, and explore wider applications of brain-inspired chips in the real world.